\UseRawInputEncoding
\documentclass[twocolumn,aps,amsmath,amssymb,color,superscriptaddress,prl,footinbib,longbibliography]{revtex4-2}
\usepackage{stix}
\usepackage[colorlinks,citecolor=blue,linkcolor=blue,urlcolor=blue]{hyperref}
\usepackage{graphicx}
\usepackage{dcolumn}
\usepackage{multirow}
\usepackage{bm}
\usepackage{times}
\usepackage{tabularx}
\usepackage{footmisc}
\usepackage{xcolor}
\usepackage{float}
\usepackage{tabularx}
\usepackage[english]{babel}
\usepackage{url}
\usepackage{hyperref}
\usepackage{ulem}
\newcommand{\red}[1]{{\leavevmode\color{black}#1}} 

\begin{document}

\title{Remote gate control of  topological transitions in moir\'{e} superlattices via cavity vacuum fields}

\author{Zuzhang Lin}
\affiliation{Department of Physics, The University of Hong Kong, Hong Kong, China}
\affiliation{HKU-UCAS Joint Institute of Theoretical and Computational Physics at Hong Kong, China}

\author{Chengxin Xiao}
\affiliation{Department of Physics, The University of Hong Kong, Hong Kong, China}
\affiliation{HKU-UCAS Joint Institute of Theoretical and Computational Physics at Hong Kong, China}

\author{Danh-Phuong Nguyen}
\affiliation{Universit\'e Paris Cit\'e, CNRS, Mat\'eriaux et Ph\'enom\`{e}nes Quantiques, 75013 Paris, France}

\author{Geva Arwas}
\affiliation{Universit\'e Paris Cit\'e, CNRS, Mat\'eriaux et Ph\'enom\`{e}nes Quantiques, 75013 Paris, France}

\author{Cristiano Ciuti}
\affiliation{Universit\'e Paris Cit\'e, CNRS, Mat\'eriaux et Ph\'enom\`{e}nes Quantiques, 75013 Paris, France}

\author{Wang Yao}
\email{wangyao@hku.hk}
\affiliation{Department of Physics, The University of Hong Kong, Hong Kong, China}
\affiliation{HKU-UCAS Joint Institute of Theoretical and Computational Physics at Hong Kong, China}

\date{\today}

\begin{abstract}
Placed in cavity resonators with three-dimensionally confined electromagnetic wave, the interaction between quasiparticles in solids can be induced by exchanging virtual cavity photons, which can have a non-local characteristic. Here we investigate the possibility of utilizing this nonlocality to realize the remote control of the topological transition in mesoscopic moir\'{e} superlattices at full filling (one electron/hole per supercell) embedded in a split-ring terahertz electromagnetic resonator. We show that gate tuning one moir\'{e} superlattice can remotely drive a topological band inversion in another moir\'{e} superlattice not in contact but embedded in the same cavity. Our study of remote on/off switching of a topological transition provides a novel paradigm for the  control of material properties via cavity vacuum fields.

\end{abstract}
\maketitle

In recent years, the strong interaction between light and condensed matter systems, typically realized in a cavity-embedded configuration,  has attracted widespread research interest \cite{garcia-vidal_manipulating_2021,sentef_cavity_2018, mazza_superradiant_2019,kiffner_mott_2019,curtis_cavity_2019,allocca_cavity_2019,latini_cavity_2019,latini_ferroelectric_2021,lenk_dynamical_2022,soykal_strong_2010,cirio2016ground,stassi2013spontaneous,di2019interaction,forn2019ultrastrong,ashida2023cavity}.
One of the most salient feature of this regime is that the exchange of virtual cavity photons can mediate a plethora of virtual-excitation-dressed ground states \cite{frisk2019ultrastrong}, including  superconductivity \cite{schlawin_cavity-mediated_2019,gao_photoinduced_2020}, superfluidity \cite{schlawin_cavity-mediated_2019-1} and charge-density-wave phases \cite{li_manipulating_2020}.
 Most importantly, the virtual-photon-mediated interaction can possess a remarkable nonlocal character when the cavity photon mode field is discretized in energy while delocalized all over the electronic sample, as it has been pioneered in the context of the quantum Hall effect \cite{ciuti2021cavity,appugliese2022breakdown,arwas2022quantum}.

 The nonlocality raises the intriguing possibility of remote control of a matter system. However, this has been largely overlooked as most research efforts regard cavity-embedded matter as a macroscopic system in the thermodynamic limit \cite{li_vacuum_2018,andolina_theory_2020,mazza_superradiant_2019,andolina_cavity_2019,latini_ferroelectric_2021}. To uncover the remote control possibilities from the nonlocal characteristic, one has to consider mesoscopic configurations. In this respect, an interesting configuration is mesoscopic moir\'{e} superlattice embedded in a metallic split-ring terahertz (THz) electromagnetic resonator \cite{scalari_ultrastrong_2012, chen_review_2016,paravicini-bagliani_gate_2017,paravicini-bagliani_magneto-transport_2019,keller_few-electron_2017,jeannin_ultrastrong_2019,appugliese2022breakdown,arwas2022quantum,ciuti2021cavity}.
The THz resonator enjoys deep subwavelength mode confinement and strongly enhanced electric field vacuum fluctuations \cite{keller_few-electron_2017,paravicini-bagliani_magneto-transport_2019}. moir\'{e} superlattice--a platform for tailoring versatile material properties--- is suitable for exploring cavity control at frequency down to the THz range, given their meV scale mini-gaps tunable by twisting angles \cite{wu2019topological}. In experimental reality, these superlattices are mesoscopic with practically limited lattice sites, having spatial dimension much smaller than the cavity mode volume of the THz resonators. Most importantly, moir\'{e} superlattices exhibit remarkable topological matter properties \cite{wu2019topological, tong2017topological,yu_giant_2020} and can serve as a prototype for remote control of topological transitions in matters.

 In this work, we demonstrate remote gate control of topological transition in a mesoscopic moir\'{e} superlattice (\red{moir\'{e} 1}), by gate tuning a second moir\'{e} superlattice (\red{moir\'{e} 2}) that shares the same cavity vacuum with \red{moir\'{e} 1}. Within a mean-field description corroborated by exact diagonalization calculations for smaller size system, we find that the presence of a moir\'{e} can perturb the cavity vacuum field, which, in turn, introduces a mass term to tune the topological transition of the moir\'{e} minibands. This forms the basis of the cavity-mediated nonlocal interaction between two moir\'{e} superlattices embedded in a common cavity. By tuning the interlayer bias applied on \red{moir\'{e} 2}, a remote control of the mini-bands Chern numbers of \red{moir\'{e} 1} can be realized, and vice versa. We emphasize that the present mechanism does not require any electronic contact between the two moir\'{e} samples which can also have different sizes and characteristic parameters. The principle can be straightforwardly extended to enable non-local interplay between multiple mesoscopic systems of distinct natures.

\begin{figure*}
\centering
\includegraphics[width=0.9\textwidth]{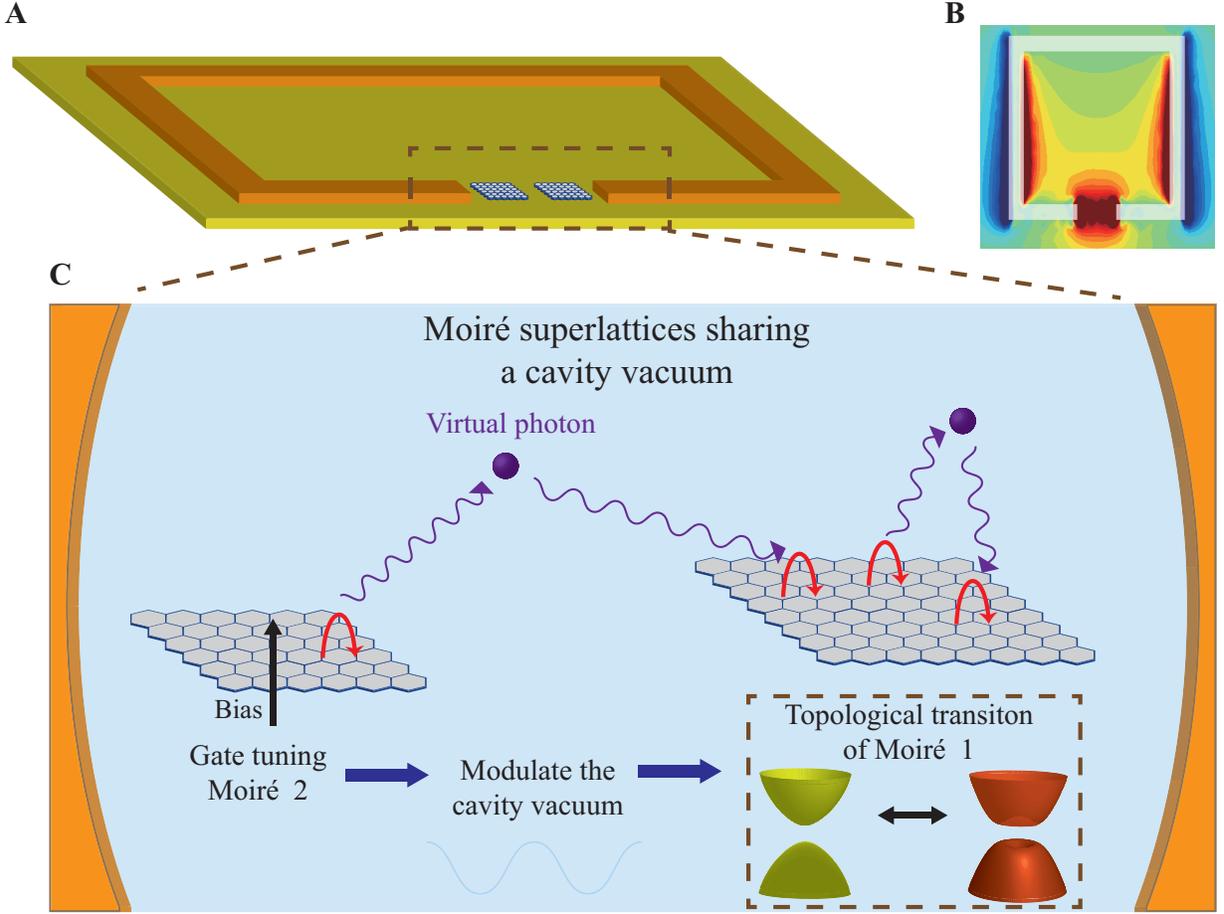}
\caption{\textbf{Sketch of the remote topological control scheme via cavity vacuum fields.} (A) Set-up with two moir\'{e} superlattices (moir\'{e} 1 and moir\'{e} 2) embedded in a metallic split-ring THz electromagnetic resonator. (B) Schematic diagram of the spatial dependence of the cavity electric field concentrated on the gap of the metallic split-ring THz electromagnetic resonator  (the red color denotes the part with the largest electric vacuum field). (C) Illustration of the physical mechanism providing  remote control of the topological transition via cavity vacuum fields. The two moir\'{e} superlattices sharing the same cavity vacuum interact by exchanging virtual photons. By gate tuning \red{moir\'{e} 2}, a topological transition is induced in \red{moir\'{e} 1}.}
\label{setup}
\end{figure*}

We consider the configuration where \red{moir\'{e} 1 and moir\'{e} 2} are embedded into a THz resonator (Fig. \ref{setup}). As an exemplary demonstration, let us assume that both moir\'{e} systems are transition metal dichalcogenides (TMDs) homobilayers with small twisting angles from 0 degree (R-type). The low energy valence states of such moir\'{e} superlattice at the $\mathbf{K}$ ($\mathbf{K^{\prime}}$) valley can be described by a two-band tight-binding (TB) model, of complex amplitude next-nearest-neighbor hopping on hexagonal superlattice sites due to the real-space Berry connection from moir\'{e} patterns, essentially a Haldane model \cite{wu2019topological,yu_giant_2020}. Without losing the essence of the physics to be discussed, we only consider one valley for each moir\'{e}, as the remote control via the cavity vacuum is valley-independent by itself. In the basis of their Bloch eigenstates, $\hat{H}_1$ and $\hat{H}_2$  can be written as $\hat{H}_1 =\sum_{\mathbf{k}} (E_{g \mathbf{k}} \hat{c}_{g \mathbf{k}}^{\dagger} \hat{c}_{g \mathbf{k}}+E_{e \mathbf{k}} \hat{c}_{e \mathbf{k}}^{\dagger} \hat{c}_{e \mathbf{k}})$ and $\hat{H}_2 =\sum_{\mathbf{q}}( \mathcal{E}_{g \mathbf{q}} \hat{d}_{g \mathbf{q}}^{\dagger} \hat{d}_{g \mathbf{q}}+\mathcal{E}_{e \mathbf{q}} \hat{d}_{e \mathbf{q}}^{\dagger} \hat{d}_{e \mathbf{q}})$, where $\hat{c}^{\dagger}$ ($\hat{c}$) and $\hat{d}^{\dagger}$ ($\hat{d}$) are creation (annihilation) fermionic operators of the quasiparticles in \red{moir\'{e} 1 and moir\'{e} 2}, respectively. Note that $E$ and $ \mathcal{E}$ are corresponding moir\'{e} mini-band energies. The subscripts $g$ and $e$ respectively refer to the lower and upper bands, while $\mathbf{k}$ and $\mathbf{q}$ are the wavevectors of \red{moir\'{e} 1 and moir\'{e} 2}, respectively. The interlayer bias applied to such moir\'{e} superlattice creates onsite energy difference between its two sublattices, which can tune the mini-bands dispersion and topology locally.

Let us now consider a single-mode cavity with a cavity field polarized along the plane of the moir\'{e} superlattices. Within the TB model of the moir\'{e} superlattices the cavity coupling is enforced via the Peierls substitution. In the following, we will consider the light-matter Hamiltonian:
\begin{equation}
\begin{aligned}
\hat{H} &=\hat{H}_1+\hat{H}_2+\hat{H}_{1 v}+\hat{H}_{2 v}+\hat{H}_f ,\\
\end{aligned}
\end{equation}
 where $\hat{H}_{f}=\hbar \omega \hat{a}^{\dagger} \hat{a}$ is the bare cavity Hamiltonian with $\omega$ the frequency of the cavity mode and $\hbar$ the reduced Planck constant. The operator $\hat{H}_{1v}=\chi\left(a+a^{\dagger}\right) \hat{\mathcal{M}}$ ($\hat{H}_{2v}=\chi\left(a+a^{\dagger}\right) \hat{\mathcal{N}}$) describes the interaction between the cavity quantized field and \red{moir\'{e} 1 (moir\'{e} 2)}, $\chi$ is the coupling strength, and $\hat{\mathcal{M}}\equiv \sum_{f i \mathbf{k}} \mathcal{M}_{f i \mathbf{k}}\hat{c}_{f \mathbf{k}}^{\dagger} \hat{c}_{i \mathbf{k}}$ ($\hat{\mathcal{N}}\equiv \sum_{f i \mathbf{q}} \mathcal{N}_{f i \mathbf{q}}\hat{d}_{f \mathbf{q}}^{\dagger} \hat{d}_{i \mathbf{q}}$) is a Hermitian operator with $i, f=g,e$. 

 The Hamiltonian $\hat{H}$ acts on a Hilbert space consisting of subspaces $\xi_n$ ($n=0,1,2...$) in which the photon number is $\langle a^{\dagger} a \rangle=n$. Following the Schrieffer-Wolff (SW) transformation \cite{schrieffer1966relation} to eliminate the light-matter interaction ($\hat{H}_{1 v}+\hat{H}_{2 v}$) to the first order, we get a block diagonalized Hamiltonian $\hat{H}_S=e^{\hat{S}}\hat{H}e^{-\hat{S}}$ with the generator $\hat{S}$ satisfying $[\hat{S},\hat{H}_1+\hat{H}_2+\hat{H}_f]=-(\hat{H}_{1 v}+\hat{H}_{2 v})$.
 Projecting the Hamiltonian $\hat{H}_S$ into the low energy sector (see Supplementary Material) gives an effective many-body Hamiltonian $\hat{H}_{\text{tot, eff }}=\hat{H}_{1, \text {eff}}+\hat{H}_{2, \text {eff}}+\hat{I}_{1}+\hat{I}_{2}$. While the first two terms  $\hat{H}_{1, \text {eff}}=\hat{H}_{1}+\frac{1}{2}[\hat{S}_1,\hat{H}_{1 v}]$ and $\hat{H}_{2, \text {eff}}=\hat{H}_{2}+\frac{1}{2}[\hat{S}_2,\hat{H}_{2 v}]$  describe the interaction of quasiparticles respectively in moir\'{e} 1 and moir\'{e} 2, the remaining two terms $\hat{I}_{1}=\frac{1}{2}[\hat{S}_1,\hat{H}_{2 v}]$ and $\hat{I}_{2}=\frac{1}{2}[\hat{S}_2,\hat{H}_{1 v}]$ denote that the quasiparticles in moir\'{e} 1 interact with quasiparticles in moir\'{e} 2. The specific form of  $\hat{I}_{1}$ is
 \begin{widetext}
\begin{equation}
\hat{I}_1=\frac{\chi^2}{2} \sum_{f i \mathbf{k}, f^{\prime} i^{\prime} \mathbf{q}}\left(\frac{\mathcal{M}_{f i \mathbf{k}} \hat{c}_{f \mathbf{k}}^{\dagger} \hat{c}_{i \mathbf{k}} \mathcal{N}_{f^{\prime} i^{\prime} \mathbf{q}} \hat{d}_{f^{\prime} \mathbf{q}}^{\dagger} \hat{d}_{i^{\prime} \mathbf{q}}}{E_{f \mathbf{k}}-E_{i \mathbf{k}}-\hbar \omega}-\frac{\mathcal{N}_{f^{\prime} i^{\prime} \mathbf{q}} \hat{d}_{f^{\prime} \mathbf{q}}^{\dagger} \hat{d}_{i^{\prime} \mathbf{q}} \mathcal{M}_{f i \mathbf{k}} \hat{c}_{f \mathbf{k}}^{\dagger} \hat{c}_{i \mathbf{k}}}{E_{f \mathbf{k}}-E_{i \mathbf{k}}+\hbar \omega}\right)
\label{Eq_I}
\end{equation}
\end{widetext}
Equation (\ref{Eq_I}) indicates that the photons absorbed (emitted) during the transitions $|i\rangle \rightarrow |f\rangle$ of particles in \red{moir\'{e} 1} will be emitted (absorbed) accompanying the transitions $|i^{\prime}\rangle \rightarrow |f^{\prime}\rangle$ of particles in \red{moir\'{e} 2}. These are the cavity-mediated interaction terms responsible for the remote control of \red{moir\'{e} 2 on moir\'{e} 1}, and vice versa.
Note that for simplicity we have omitted the Coulomb electron-electron interaction terms in each moir\'{e} superlattice assuming that Coulomb interaction is strongly screened by a dielectric substrate. Regardless, the remote control scheme would remain the same with the cavity-mediated interaction between remote moir\'{e} superlattices.

\begin{figure}
\centering
\includegraphics[width=1\columnwidth]{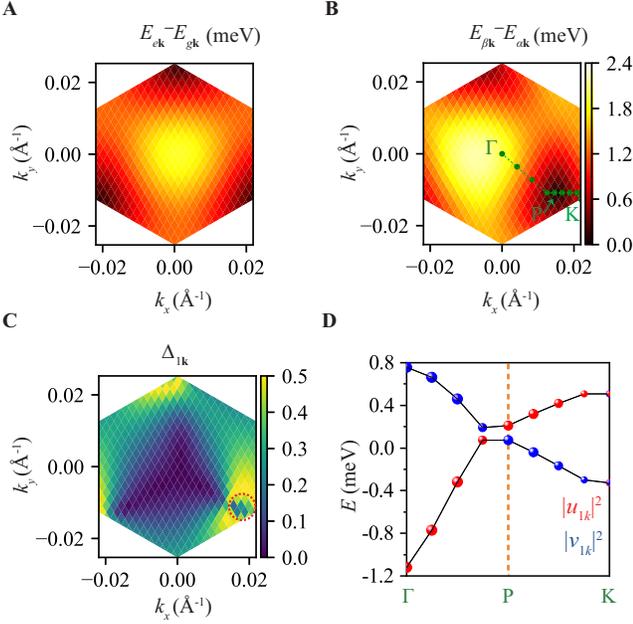}
\caption{\textbf{Inter-miniband transition energy, mean-field order parameter, and topological band inversion.} (A-B) Color plot of the inter-miniband transition energy in reciprocal space for \red{moir\'{e} 1} consisting of \red{21 by 21} superlattice sites. Calculation in (A) uses the bare Hamiltonian $\hat{H}_1$, i.e. in the absence of cavity, and (B) uses $\hat{H}_{1,\text{MF}}$ in the presence of cavity quantized field perturbed by a second superlattice (\red{moir\'{e} 2}). See text. (C) Color plot of the mean-field order parameter \red{$\Delta_{1 \mathbf{k}}$} in reciprocal space, corresponding to the calculation in (B). (D) The miniband dispersion predicted by moir\'{e} 1's mean-field Hamiltonian $\hat{H}_{1,\mathrm{MF}}$. The eigenstate amplitudes on to moir\'{e} 1's bare Hamiltonian basis are indicated by the size of the spheres.}.
\label{Fig_Band}
\end{figure}

\begin{figure*}
\centering
\includegraphics[width=0.9\textwidth]{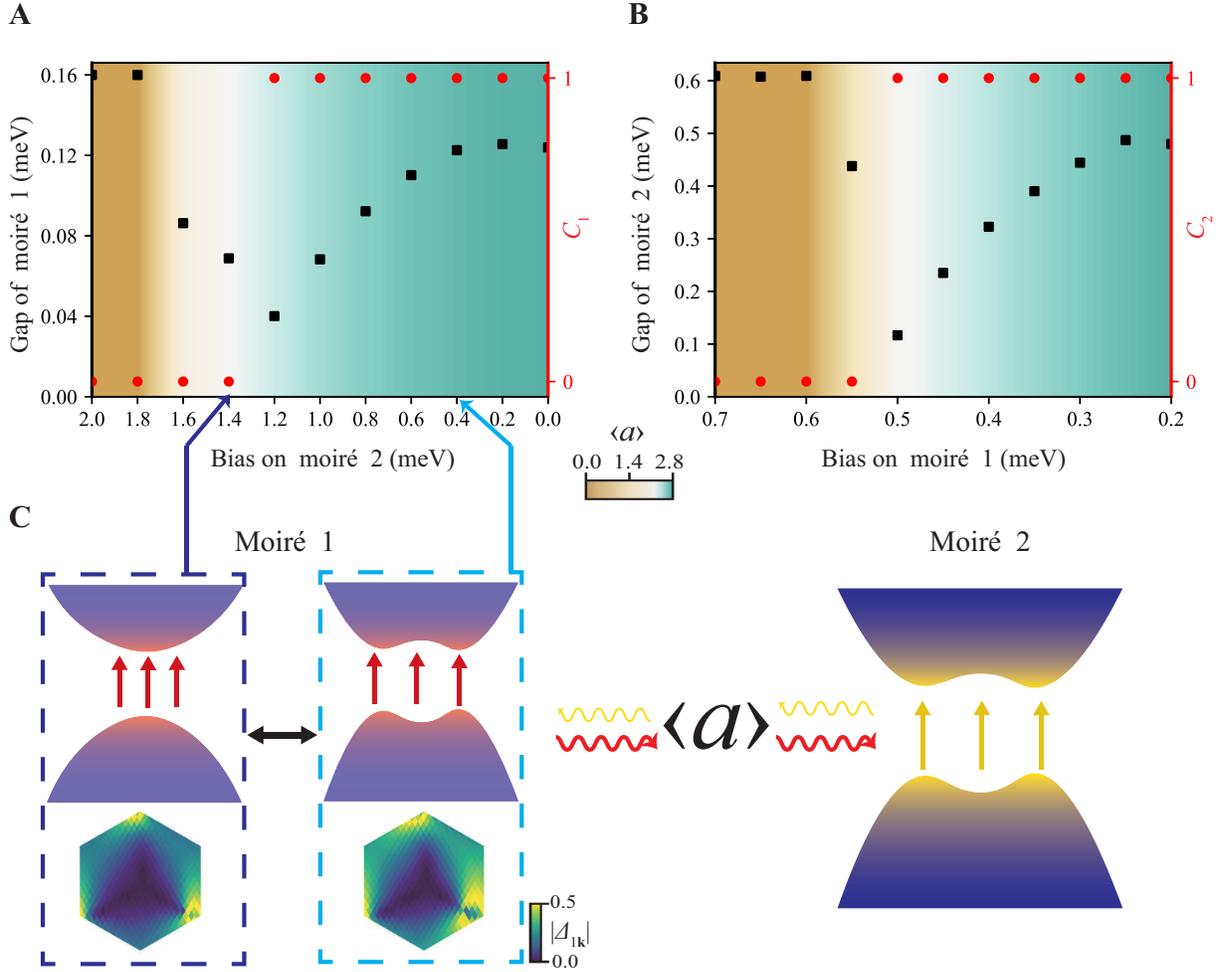}
\caption{
\textbf{Remote gate control of topological transitions.} (A) Topological transition of \red{moir\'{e} 1} controlled remotely by the interlayer bias of \red{moir\'{e} 2 ($V_2$)}, while fixing \red{moir\'{e} 1's} own bias at \red{0.7 meV}. The change of cavity field $\langle a \rangle$, \red{moir\'{e} 1's} Chern number and gap as function of \red{$V_2$} are shown respectively by the background color, red dots, black \red{squares}.  (B) The reciprocal topological control of \red{moir\'{e} 2} remotely by the interlayer bias of \red{moir\'{e} 1 ($V_1$)}, while fixing \red{moir\'{e} 2's bias at 2.2 meV}. (C) Schematic diagram of the interaction due to the exchange of virtual photons between moir\'{e} 1 and moir\'{e} 2. At two \red{$V_2$} values, the mean-field order parameter \red{$\Delta_{1\mathbf{k}}$} is shown, where moir\'{e} 1 is topologically trivial and nontrivial respectively.}
\label{FIG3}
\end{figure*}

Within a mean-field framework, we can approximate the bilinear terms in $\hat{H}_{\text{tot, eff }}$ as $\hat{O} \hat{O}^{\prime} \approx\langle O\rangle \hat{O}^{\prime}+\hat{O}\langle O^{\prime}\rangle-\langle O \rangle \langle O^{\prime}\rangle$, where $\hat{O}, \hat{O}^{\prime}=\hat{c}_{f \mathbf{k}}^{\dagger} \hat{c}_{i \mathbf{k}}, \hat{d}_{f \mathbf{q}}^{\dagger} \hat{d}_{i \mathbf{q}}$. By grouping the resulting terms according to the operators $\hat{c}_{f \mathbf{k}}^{\dagger} \hat{c}_{i \mathbf{k}}$ and $\hat{d}_{f \mathbf{q}}^{\dagger} \hat{d}_{i \mathbf{q}}$, we get a mean-field Hamiltonian $\hat{H}_{\mathrm{MF}}=\hat{H}_{1,\mathrm{MF}}+\hat{H}_{2,\mathrm{MF}}$, where $\hat{H}_{1,\mathrm{MF}}$ and $\hat{H}_{2,\mathrm{MF}}$ respectively describe the mean-field effects on \red{moir\'{e} 1 and moir\'{e} 2}:
\begin{equation}
\begin{aligned}
\hat{H}_{1,\mathrm{MF}}=&\sum_{\mathbf{k}}[\tilde{E}_{g \mathbf{k}} \hat{c}_{g \mathbf{k}}^{\dagger} \hat{c}_{g \mathbf{k}}+\tilde{E}_{e \mathbf{k}} \hat{c}_{e \mathbf{k}}^{\dagger} \hat{c}_{e \mathbf{k}}
+(\tilde{t}_{1\mathbf{k}} \hat{c}_{g \mathbf{k}}^{\dagger} \hat{c}_{e \mathbf{k}}+h . c .)]+\varepsilon_{1} \, ,\\
\hat{H}_{2,\mathrm{MF}}=&\sum_{\mathbf{q}}[\tilde{\mathcal{E}}_{g \mathbf{q}} \hat{d}_{g \mathbf{q}}^{\dagger} \hat{d}_{g \mathbf{q}}+\tilde{\mathcal{E}}_{e \mathbf{q}} \hat{d}_{e \mathbf{q}}^{\dagger} \hat{d}_{e \mathbf{q}}+(\tilde{t}_{2\mathbf{q}} \hat{d}_{g \mathbf{q}}^{\dagger} \hat{d}_{e \mathbf{q}}+h . c .)]+\varepsilon_{2}.
\end{aligned}
\end{equation}
Here $\tilde{E}_{g \mathbf{k}}$, $\tilde{E}_{e \mathbf{k}} $, $\tilde{t}_{1 \mathbf{k}}$, $\tilde{\mathcal{E}}_{g \mathbf{q}}$, $\tilde{\mathcal{E}}_{e \mathbf{q}} $, $\tilde{t}_{2 \mathbf{q}}$, $\varepsilon_{1}$ and $\varepsilon_{2}$ are  renormalized parameters with mean-field corrections (see details in Supplementary Material).  The many-body ground state $
|\Psi\rangle=\prod_{\mathbf{k}}\left(u_{1 \mathbf{k}}^* \hat{c}_{g \mathbf{k}}^{\dagger}+v_{1 \mathbf{k}}^* \hat{c}_{e \mathbf{k}}^{\dagger}\right) \prod_{\mathbf{q}}\left(u_{2 \mathbf{q}}^* \hat{d}_{g \mathbf{q}}^{\dagger}+v_{2 \mathbf{q}}^* \hat{d}_{e \mathbf{q}}^{\dagger}\right)|0\rangle
$ features interband coherence of \red{moir\'{e} 1 (moir\'{e} 2)} characterized by the mean-field order parameter $\Delta_{1{\mathbf{k}}}=v_{1\mathbf{k}}u^{*}_{1\mathbf{k}}$ ($\Delta_{2{\mathbf{q}}}=v_{2\mathbf{q}}u^{*}_{2\mathbf{q}}$).  The mean-field order parameters can be solved self-consistently through the following gap-like equations
\begin{equation}
\begin{aligned}
\Delta_{1{\mathbf{k}}}&=-\frac{\tilde{t}_{1\mathbf{k}}}{\sqrt{4\left|\tilde{t}_{1\mathbf{k}}\right|^{2}+\left(\tilde{E}_{e \mathbf{k}}-\tilde{E}_{g \mathbf{k}}\right)^{2}}}, \\
\Delta_{2{\mathbf{q}}}&=-\frac{\tilde{t}_{2\mathbf{q}}}{\sqrt{4\left|\tilde{t}_{2\mathbf{q}}\right|^{2}+\left(\tilde{\mathcal{E}}_{e \mathbf{q}}-\tilde{\mathcal{E}}_{g \mathbf{q}}\right)^{2}}}. \label{Gap}
\end{aligned}
\end{equation}
Note that these two equations are not independent: the order parameters of \red{moir\'{e} 1} are affected by the order parameters of \red{moir\'{e} 2}, and vice versa. To test the validity of our mean-field approach, we have also performed exact diagonalization results with a small number of electrons, yielding the same qualitative results (see the Supplementary Material).

In the calculations presented below, to exemplify the dissimilarities of the two moir\'{e} superlattices, we use a \red{21 by 21} superlattice for moir\'{e} 1 with strengths of the nearest and next-nearest neighbor hopping being \red{0.29 meV and 0.06 meV} respectively \cite{wu2019topological}, while moir\'{e} 2 is a \red{10 by 10} superlattice with the corresponding hopping amplitudes being 0.5 meV and 0.2 meV instead \cite{yu_giant_2020}. The phase of the next-nearest-hoping is $2\pi/3$ for both moir\'{e}s, corresponding to a \red{positive} flux Haldane model from valley \red{K}. We consider a THz resonator cavity mode of volume \red{$V=7 \times 10^{6}$ nm$^3$} and \red{quantized mode energy $\hbar \omega =$ 8.1 meV}, which leads to a light-matter coupling strength $\chi=\red{0.17}$. More details are given in Supplementary Material.

As an example, we first solve the gap equation by fixing  the interlayer bias applied on moir\'{e} 1 at  \red{0.7} meV. In the absence of the cavity quantized field (i.e., $\chi=0$), moir\'{e} 1 displays an electronic band gap at the $\mathbf{K}$ point (Fig. \ref{Fig_Band}A) and is topologically trivial. Embedding moir\'{e} 1 {\it alone} in the cavity, we find negligible change is introduced to its electronic structure at the given bias, whereas the cavity vacuum is also negligibly perturbed. When the cavity also hosts a second moir\'{e}, tuning the bias of moir\'{e} 2 can drastically change the electronic structure of moir\'{e} 1. In Fig. \ref{Fig_Band}B, we plot the mini-band transition energies of moir\'{e} 1, calculated when moir\'{e} 2 is biased at \red{0.2} meV,  which now exhibits an electronic band gap at the $\mathbf{P}$ point instead. Its ground state has a pronounced interband coherence near the $\mathbf{K}$ point (Fig. \ref{Fig_Band}C), which is reasonable according to the Eq. (\ref{Gap}) as $\tilde{\mathcal{E}}_{e \mathbf{k}}-\tilde{\mathcal{E}}_{g \mathbf{k}} \propto \mathcal{E}_{e \mathbf{k}}-\mathcal{E}_{g \mathbf{k}} $ reaches the smallest value near the $\mathbf{K}$ point.

Notably, a small circular region (indicated by the red dashed circle in Fig. \ref{Fig_Band}C) where \red{$\Delta_{1\mathbf{k}}$} is almost zero, is surrounded by the areas with maximal interband coherence \red{($\Delta_{1\mathbf{k}} \sim 0.5$)}. We note that the zero \red{$\Delta_{1\mathbf{k}}$} at this region is due to band inversion \red{($v_{1\mathbf{k}}=1, u_{1\mathbf{k}}=0$)}, which is different from the zero \red{$\Delta_{1\mathbf{k}}$} elsewhere (e.g. in the region near $\Gamma$ point where \red{$v_{1\mathbf{k}}=0, u_{1\mathbf{k}}=1$}). This is confirmed by the band dispersion of the Hamiltonian \red{$\hat{H}_{1,\mathrm{MF}}$} and by the wavefunction projections on the original Hamiltonian basis (Fig. \ref{Fig_Band}D). Furthermore, the calculated Chern numbers of the lower and upper bands are found to be 1 and -1, respectively. Therefore, in the presence of moir\'{e} 2, the cavity-mediated interaction has provided a topological nontrivial mass term to moir\'{e} 1.

This topological nontrivial mass term on \red{moir\'{e} 1} arising from the cavity-mediated coupling is tunable by the interlayer bias on moir\'{e} 2. As a result, gate tuning \red{moir\'{e} 2} will realize a remote control of the topological transition in \red{moir\'{e} 1}. By tuning the interlayer bias on \red{moir\'{e} 2} (denoted as $V_2$ hereafter) from \red{2 meV} to \red{0 meV}, we indeed observe the gap of \red{moir\'{e} 1} closes and reopens at a critical value of \red{1.2 meV} of $V_2$, accompanied by a corresponding step change in Chern number from zero to one (Fig. \ref{FIG3}A). Conversely, the remote control of topological transition in \red{moir\'{e} 2} by gate tuning \red{moir\'{e} 1} ($V_1$) can also be realized (Fig. \ref{FIG3}B).

To reveal the physical insight of the remote control, we calculate the expectation value of field operator $\hat{a}$ to the leading order
\begin{equation}
\begin{aligned}\label{field_a}
\langle  a \rangle & =\langle \Psi | e^{\hat{S}} \hat{a} e^{-\hat{S}} | \Psi \rangle \\
  & \sum_{f i \mathbf{k}} \frac{\chi \mathcal{M}_{f i \mathbf{k}}\left\langle\hat{c}_{f \mathbf{k}} \hat{c}_{i \mathbf{k}}\right\rangle}{E_{i \mathbf{k}}-E_{f \mathbf{k}}-\hbar \omega}+\sum_{f i \mathbf{q}} \frac{\chi \mathcal{N}_{f i \mathbf{q}}\left\langle\hat{d}_{f \mathbf{q}} \hat{d}_{i \mathbf{q}}\right\rangle}{\mathcal{E}_{i \mathbf{q}}-\mathcal{E}_{f \mathbf{q}}-\hbar \omega} \\
\end{aligned}
\end{equation}
as a function of the interlayer bias. As shown in Fig. \ref{FIG3}A,  when \red{$V_2$} is tuned from \red{2 meV to 1.8 meV}, $\langle  a \rangle$ is negligibly small and the gap of moir\'{e} 1 remains  unchanged. Further reducing $V_2$, $\langle  a \rangle$ starts to increase noticeably, and at the same time the gap of moir\'{e} 1 starts to change, and eventually a topological transition occurs. Therefore, the remote control of topological transition in \red{moir\'{e} 1} is realized through modulating the cavity vacuum upon gate tuning \red{moir\'{e} 2.}
We find that nonzero $\langle  a \rangle$
 \footnote{The nonzero mean values of the field operator $\hat{a}$ does not imply a macroscopic photon condensation prevented by the no-go theorem \cite{andolina_cavity_2019, guerci_superradiant_2020,rzazewski_phase_1975,bialynicki-birula_no-go_1979,gawedzki_no-go_1981}. The moir\'{e} superlattices embedded in a metallic split-ring THz resonator form a mesoscopic system, where both the superlattice size and cavity volume are finite. In the literature convention \cite{li_vacuum_2018,andolina_theory_2020,mazza_superradiant_2019,andolina_cavity_2019,latini_ferroelectric_2021}, the field amplitude $A_0$ $\propto 1/\sqrt{\mathcal{V}}$ ($\mathcal{V}$ the mode volume of the cavity photon) is infinitesimal in the thermodynamic limit in which $\mathcal{V}$ is assigned to be infinite.}
 occurs simultaneously with the interband coherence of the electronic many-body ground state. In parameter regimes where $\langle  a \rangle$ vanishes, both moir\'{e}s have negligible interband coherence in the ground states and have no response to the remote control gate. The threshold $\langle a \rangle$ value needed to bring a moir\'{e} across the topological transition point depends on the parameters of its bare Hamiltonian without the cavity (c.f. Fig. \ref{FIG3}A and \ref{FIG3}B).

 We also notice that the light-matter coupling terms $\chi(\hat{a}^{\dagger}+\hat{a})\sum_{\mathbf{k}} \mathcal{M}_{g g \mathbf{k}}\hat{c}_{g \mathbf{k}}^{\dagger} \hat{c}_{g \mathbf{k}}$ and $\chi(\hat{a}^{\dagger}+\hat{a})\sum_{\mathbf{q}} \mathcal{N}_{g g \mathbf{q}}\hat{d}_{g \mathbf{q}}^{\dagger} \hat{d}_{g \mathbf{q}}$, which perturb the cavity vacuum while leaving the electronic state unaffected, are essential here. The nonzero value of  $ \mathcal{M}_{g g \mathbf{k}}$ and $\mathcal{N}_{g g \mathbf{q}}$ are allowed here by the lack of parity in eigenstates of Hamiltonians $\hat{H}_1$ and $\hat{H}_2$ as the out-of-plane mirror symmetry is broken in twisted TMDs bilayers. The expectation values
$\langle\mathcal{M}_{g g \mathbf{k}}\hat{c}_{g \mathbf{k}}^{\dagger} \hat{c}_{g \mathbf{k}}\rangle$ and $\langle\sum_{\mathbf{q}} \mathcal{N}_{g g \mathbf{q}}\hat{d}_{g \mathbf{q}}^{\dagger} \hat{d}_{g \mathbf{q}}\rangle$ vanish in the ground states of the bare moir\'{e} Hamiltonians $\hat{H}_1$ and $\hat{H}_2$ respectively, but become finite in the ground states of their mean-field interaction Hamiltonian $\hat{H}_{1, \text{MF}}$ and $\hat{H}_{2, \text{MF}}$ under bias parameters where interband coherence spontaneously emerges.

 In conclusion, we have shown that by gate tuning a remote moir\'e superlattice it is possible to induce a topological transition  of a second mesoscopic moir\'e system via cavity vacuum field. The remote cascade control of multiple moir\'{e} superlattices embedded in one cavity is possible following the same scheme. Besides topological transitions, the mesoscopic system consisting of cavity-embedded moir\'{e} superlattices may also provide an exciting platform to investigate the possible remote control of other physical properties, such as superconductivity and ferromagnetism.

\textbf{Acknowledgment:} We thank Hsun-Chi Chan for helping us generate Fig. 1B. \textbf{Fundings:} The work is  supported by Research Grant Council (RGC) of Hong Kong SAR China through grants HKU SRFS2122-7S05 and AoE/P-701/20, and a grant under the ANR/RGC Joint Research Scheme sponsored by RGC and French National Research Agency (A-HKU705/21, ANR-21-CE30-0056-01). W.Y. also acknowledges support by Tencent Foundation. G.A. and C.C. also acknowledge support from the Israeli Council for Higher Education - VATAT.  \textbf{Data and materials availability}: All data needed to evaluate the conclusions in the paper are present in the paper and/or the Supplementary Materials. All data related to this study may be available from the corresponding author upon reasonable request.

\bibliography{cavity_mediated_interaction}

\end{document}